# ITLingo Research Initiative in 2022

**Alberto Rodrigues da Silva**

INESC-ID, Instituto Superior Técnico, Universidade de Lisboa, Portugal

Several surveys and studies have noticed that cost, and quality problems result from mistakes that occurred in the early phases of the projects, for instance: poor definition of the project vision and respective value for the organization; misalignment between IT and business resources; failures in project management practices; or poor and low-quality technical documentation. These facts have emphasized the need for improving socio-technical disciplines such as project management, enterprise architecture, requirements engineering, or system design. These studies also noted the need to reduce the efforts involved in traditional development processes, for example, by automating some human-intensive and error-prone tasks.

This article presents the scientific project we have conducted in these last years named "ITLingo". ITLingo is a research initiative that has proposed new languages, tools, and techniques to support users to improve such practices, mainly related to those disciplines. ITLingo users are IT engineers and managers in multiple roles like project managers, enterprise architects, business analysts, system architects, requirements engineers, or even system developers. The article describes the innovative nature of the activities carried out under the ITLingo umbrella and identifies future and open challenges.

Section 1 introduces the general context, problems, and challenges of the area, Section 2 describes the scientific aspects and experiences that have been developed, and Section 3 presents a discussion and identifies open issues to be considered in future work.

## 1. Introduction

The impact and rapid evolution of information technologies (IT) have introduced challenges to organizations and society. These technologies have boosted the emergence of new industries, reinforced the importance of new businesses, and eliminated others. This situation has shown the relevance of computer engineering and the increasing awareness that it integrates and crosses new knowledge and skills, such as project management, communication, and team leadership, with technical and traditional areas like design, implementation, verification, and validation of systems. This situation has also shown that all activities with higher quality and more efficiency shall be carried out.

Our research has been mainly directed toward studying the best practices in computer engineering and their application in the domain of information systems (IS). An IS is considered a complex system that integrates both human and technological resources that make it possible to satisfy the information needs of organizations and their business processes [Bri2000, Dav2020]. ISs are considered essential to support implementation or reengineering strategies and to obtain competitive advantages – with impact in terms of cost reduction, time to market, or service innovation – and are commonly associated with digital transformation projects [Bri2000, Alt2008].

After identifying the ISs (or computer-based systems) necessary to develop and integrate within an organization, their implementation is commonly considered by the software engineering area [Som2018, Pre2019]. The activities associated with software engineering can be grouped into three main phases – design, implementation, operation, and maintenance –, supported by elementary disciplines, such as requirements engineering, software design, programming, or software quality. Throughout each generic phase, concrete activities are executed that produce specific deliverables in planned milestones. It also integrates a set of transversal project management activities (specifically scope, time, cost, quality, etc.). Some of these disciplines (e.g., requirements engineering, project management, programming, software quality) are presented and discussed in the scope of international computing curricula (such as SE2014 [SE2014], IS 2020 [IS2020]), bodies of knowledge (e.g., SWEBOK [BF2014], PMBOK [PMI2021] or BABOK [IIBA2015]), or popular pedagogical books [RR2006, Poh2010, Sch2011, CB2020, Kot2021.

However, several studies have steadily reported the low success rate of IT projects, which are mostly cancelled, finished but exceeding the initial deadline or budget, or changing the original scope. There are some causes reported as contributing to these results, namely [EV2010, Sta2016, Har2018, KW2018, RZ2020]: (c1) lack of commitment from the executives; (c2) lack of commitment and engagement from users; (c3) misunderstanding the business value of IT systems; (c4) poor communication among stakeholders; (c5) failures in project management; (c6) inability to manage requirements; (c7) poor software quality; (c8) shortage of human resources with the appropriate competencies and skills; (c9) organizational politics and change.

Some of these causes (e.g., c2 to c7) are mainly related to the challenges addressed in the context of our research, namely in what concerns the need for (1) the adoption of more rigorous and consistent technical specifications and models, which would enable the sharing of a shared and clear vision of the system to be built; (2) better communication and coordination between project stakeholders; and (3) modularity and reuse mechanisms that would contribute to increasing the overall quality and productivity of these processes.

## 2. ITLingo Initiative

This section presents the research initiatives we have conducted over the last years related to the abovementioned problems and challenges. These initiatives were an umbrella context of several projects that attracted dozens of researchers and students, integrated initially into the "Information Systems Group" (GSI[1]) and then in the scientific area "Information and Decision Support Systems" (IDSS[2]) of INESC-ID.

This section introduces two former research initiatives and then presents the ITLingo, which is the initiative that covers our ongoing research areas.

### 2.1. Former Research

Roughly between 2004 and 2017, we conducted several research initiatives and projects at INESC-ID; from those, I would like to underline two – the ProjectIT and the MDDLingo initiatives – mainly related to ITLingo.

**ProjectIT**[3] **(2004 to 2013)** was a research initiative focused on the design, implementation, and project management of information systems. In the original definition of this initiative, a set of guiding principles was considered, some focusing on organizational aspects and others with a more technical nature. Among these principles, we may recall the following [Sil2004]: projects shall be aligned with and driven by the business; involve customers and users throughout the project lifecycle; adopt agile project management practices; support the communication with visual models, or adopt model-based software development approaches. This research was developed through several experiences around two complementary areas: ProjectIT/Processes&Projects and ProjectIT/Requirements&Models, which were validated with the implementation of some prototypes such as the ProjectIT-Enterprise and the ProjectIT-Studio tools. ProjectIT-Enterprise was a collaborative web platform that supported the collaboration of teams focused on activities such as organization definition, process definition, project management, document management, and alignment between projects and processes [MS2010]. On the other hand, ProjectIT-Studio was a desktop tool in which we extensively researched the tight integration between model-driven and requirements engineering techniques [SS+2007, Sil2015] and introduced our initial research on modeling languages like XIS [SS+2007a] for the design of interactive applications, i.e., business information systems, or specification language for requirements engineering, with the RSL-IL [DS2013].

---

[1] GSI web site: http://isg.inesc-id.pt

[2] IDSS web site: http://idss.inesc-id.pt/

[3] ProjectIT web site: http://isg.inesc-id.pt/alb/ProjectIT

**MDDLingo[4] initiative (2013 to 2017)** was focused on model-driven engineering approaches (MDE) [Voe+2013, Sil2015] and aimed to investigate techniques to increase the abstraction and productivity of the design and development of information systems. We faced these challenges by creating new domain-specific languages – i.e., languages with constructs closer to the domain and not to the technology concepts – that make them easier and faster to be used [MHS2005]. In the scope of this initiative, the implementation and use of modeling languages were extensively explored to help IT engineers design and analyze computer applications for different domains and types of technologies. For instance, we defined languages (with respective tool support) such as (i) XIS-Mobile for the design of mobile applications in a platform-independent approach [RS2014]; (ii) XIS-Web for the design and rapid prototyping of multi-device web applications [SRS2019]; or (iii) DSL3S for the design of spatial simulation scenarios [SS2015]. These languages were defined and implemented with meta-modeling techniques as UML profiles [MHS2005, Voe+2013, RSS2016]. The investigation results were validated with case studies in different application fields, some with industry partners.

## 2.2. Current Research

The ITLingo research initiative intends to improve the quality and efficiency of how engineers and domain experts produce and manage technical documentation in the IT domain. ITLingo has the following specific objectives that are further discussed below: (O1) Research and develop specification languages to support diverse IT disciplines, such as requirements engineering or project management. (O2) Research and discuss linguistic patterns, styles, and guidelines to help write better technical documentation. (O3) Research the automation of human-intensive tasks in producing technical documentation and related artefacts, such as text extraction, document validation, document automation, or even the generation of software applications from platform-independent specifications. (O4) Research and implement computational tools and platforms to demonstrate the proposed ideas, relevance, and usefulness. (O5) Finally, apply and validate these results with concrete case studies in collaboration with the industry and its professionals and produce reusable libraries of technical specifications.

**Specification Languages**

We conducted some projects related to the design and implementation of modeling or specification languages, such as (1) the modeling languages referred above in the context of MDDLingo initiative; (2) the RSL-IL language [FS2012, FS2013], which was an initial attempt of a requirements specification language; (3) the RSL-IL4Privacy [CR2015, Car+2019], for the specification of privacy policies; and (4) the SilabMDD [Sav+2015, Sav+2015a] for the specification of information system requirements based on use cases.

Based on these experiences, we have defined some languages to improve the writing of technical documentation, namely for the following disciplines: (1) project management with PSL (Project Specification Language); (2) enterprise architecture with EASL (Enterprise Architecture Specification Language); (3) requirements and tests engineering with RSL (Requirements and Tests Specification Language); and (4) application design with ASL (Application Specification Language). These are called "specification languages" or "controlled natural languages", considering they have a familiar syntax close to natural language. Their specifications can be read and understood by people with no technical background, but they can still be maintained and processed automatically by software tools. Throughout this document, we use the term "*SL" to denote any of these languages, i.e., PSL, AESL, RSL, or ASL.

---

[4] MDDLingo GitHub: https://github.com/MDDLingo

Figure 1 illustrates the evolution of these languages over the last years and their influence dependencies. For example, the first version of the RSL language was defined initially in 2016, and its design was influenced by RSL-IL [FS2012, FS2013], XIS* [RS2014, RS2018, SRS2019], and RSL-IL4Privacy [Car+2019]. However, RSL has evolved since then, for instance, by supporting several types of requirements (e.g., goals, use cases, user stories, quality requirements) [Sil2019], by integrating requirements with tests [MPS2019, PMS2020, GPS2021], or by supporting a rigorous specification of data entities and concrete data [SS2021]. On the other hand, the design of each language has influenced the creation of the others; thus, they share common aspects such as syntax and architecture. For example, the definition of the DataEntity and related elements have a common structure and syntax in both RSL and ASL languages [Sil2019, GS2020].

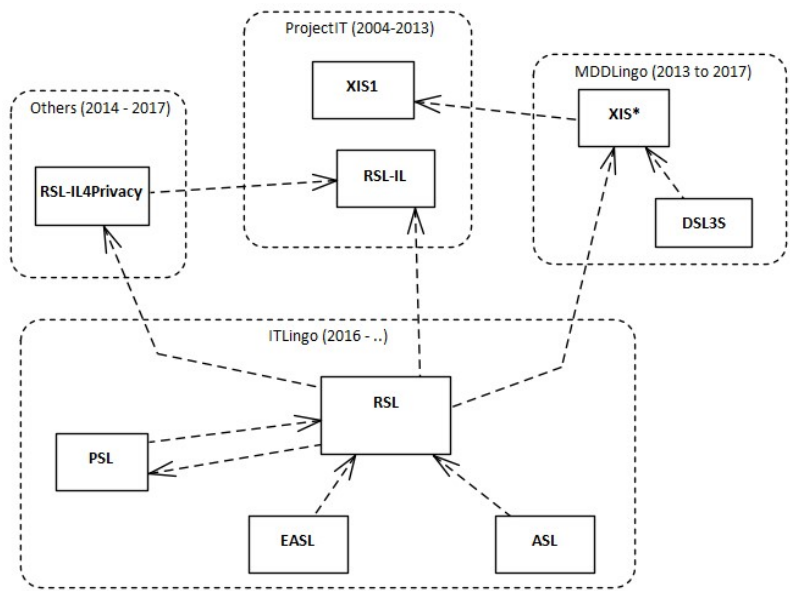

**Fig. 1**. Historical view of the research initiatives with their proposed languages (UML notation).

## Software Tools

Conceptually *SLs are process- and tool-independent languages, meaning they can be used and adopted by various users and organizations with different methods and supported by other software tools. Furthermore, *SL languages can be implemented with different technologies and multiple representations depending on the required formality, such as tabular, graphical, or textual representations [Erd+2013, Car+2019].

However, in practice, *SLs have been implemented with the Xtext[5] framework in an Eclipse-based tool called "ITLingo-Studio". With this tool, the *SL specifications can be written rigorously, automatically validated, and transformed into multiple representations and formats. Complementary, some lightweight and easy-to-use tools have been provided as MS-Excel templates; for example, see former versions of the RSL and PSL Excel templates publicly available at github[6,7]. As current work, we have developed a web version of the ITLingo-Studio, named "ITLingo-Cloud". This tool is a multi-organization and multi-project collaborative platform. Its users would manage technical documentation in an easy-to-use and more effective way than the desktop Studio version.

---

[5] Xtext: https://eclipse.org/Xtext/
[6] RSLExcel template (former version): https://github.com/ITLingo/RSL
[7] PSLExcel template (former version): https://github.com/ITLingo/PSL

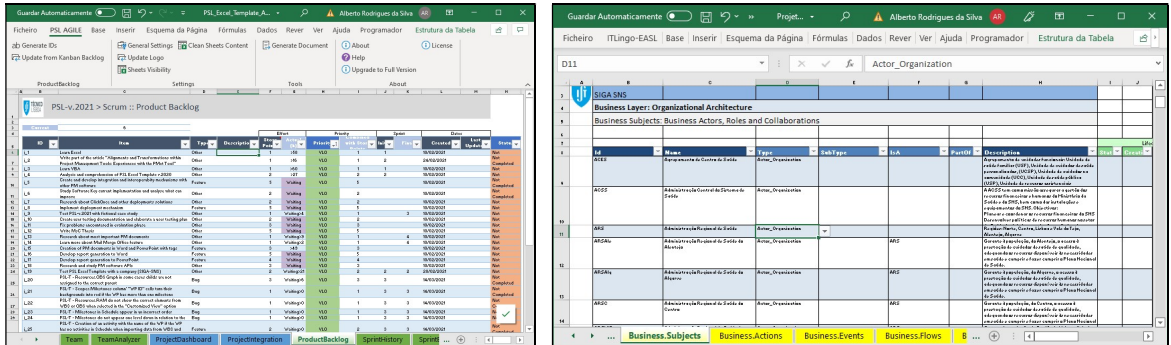

**Fig. 2**. Screenshots of the PSL-Agile (left) and EASL (right) Excel templates.

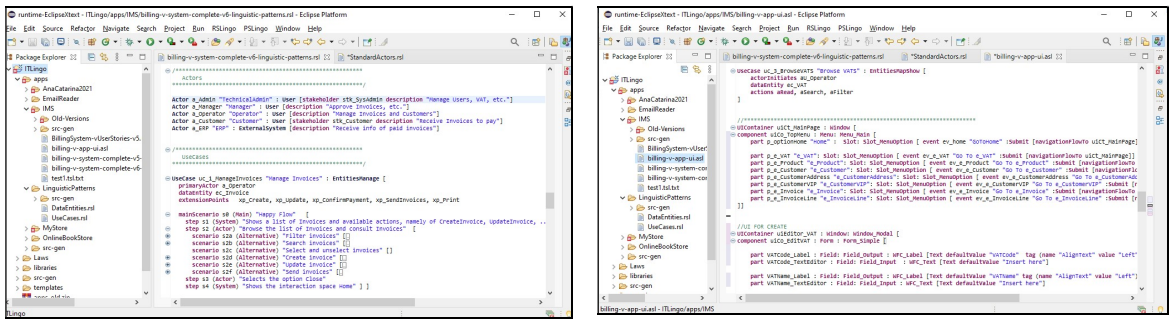

**Fig. 3**. Screenshots of the ITLingo-Studio with the authoring of RSL (left) and ASL (right image) specifications.

**\*SL Excel Template.** An \*SL MS-Excel template includes predefined sheets with data tables, which its users can quickly fill and manage. For example, the PSL-Agile Excel template contains a specific sheet with the product backlog, another with the current spring backlog, etc. These templates have analysis reports and graphics on top of that information and provide simple validation and document automation features. Figure 2 illustrates screenshots of the PSL-Agile and EASL Excel templates.

**ITLingo-Studio.** ITLingo-Studio's textual editor allows creating and editing \*SL files as provided to programming languages. This tool is developed with the Xtext framework, which, from a grammar definition is possible to generate a complete language infrastructure (e.g., parser and type-checker) and a fully customizable Eclipse plugin containing the \*SL editor with helpful features like syntax highlighting, error checking, auto-completion, or source-code navigation. Xtext-based languages have Ecore as metamodel. Since Xtext relies on EMF[8], it can be combined with other popular Eclipse plugins[9], like Xtend, Sirius, or Acceleo. Figure 3 illustrates screenshots of the ITLingo-Studio.

**ITLingo-Cloud.** ITLingo-Cloud would be designed and developed during the following years to provide a fast-to-setup and easy-to-use collaborative authoring environment for these \*SL languages. ITLingo-Cloud is being designed as a multi-organization and multi-project cloud IDE platform, inspired by emerging solutions like Google Cloud Shell Editor, AWS Cloud9, and Github Codespaces, and considering Cloud IDE frameworks like Eclipse Theia or Eclipse Che because they support Xtext-based plugins. ITLingo-Cloud shall support the writing and production of technical documentation based on the \*SL languages introduced above, as well as automation tasks discussed below, such as automatic text extraction, automatic validation, or document automation.

---

[8] EMF (Eclipse Modeling Framework): https://www.eclipse.org/modeling/emf/

[9] Eclipse market place: https://marketplace.eclipse.org/

## Linguistic Patterns, Styles, and Guidelines

As discussed initially in [Sil2017], a *linguistic pattern* is a set of rules that defines the elements and vocabulary used in the sentences of technical documents. An *element rule* consists of a group of element attributes. A *vocabulary rule* defines a set of literal terms used to categorize some element attributes and to restrict the use of a limited number of terms. As suggested in Figure 4, a *linguistic style is a concrete representation of a linguistic pattern*, which means that a linguistic style is a specific template to which attributes of the linguistic pattern can be replaced. A *linguistic pattern can be represented by multiple linguistic styles,* as discussed in recent papers [Sil2017, SS2021, Sil2021]. Furthermore, *practical guidelines* help users write more systematically and consistently, as, for instance, we did in a recent paper focused on writing requirements based on use cases and structured scenarios [Sil2021]. For future work, we intend to research more linguistic patterns, for example, for RSL: user stories, goals, and other types of requirements, as well as for test cases and acceptance criteria. We also intend to explore different writing styles, including combining textual with tabular and visual representations, and then implement document automation features, namely generating concrete documents based on these writing styles from intermediate formats, like those defined with *SLs languages.

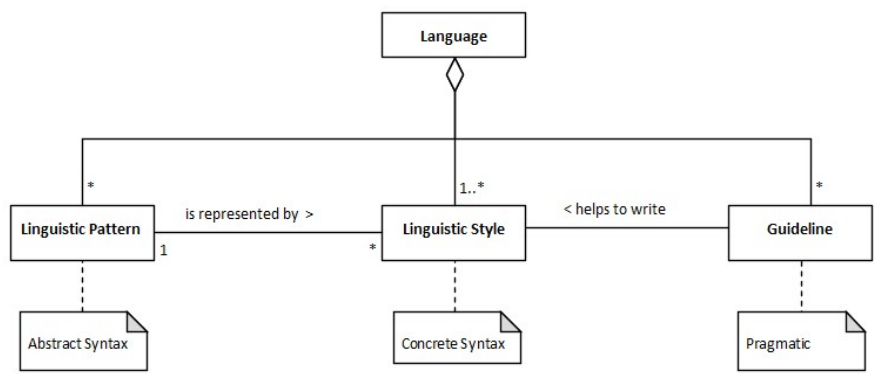

**Fig.4**. Relation between linguistic patterns, styles, and guidelines (UML notation).

## Tasks Automation

The tools introduced above shall include a set of features to allow the automation of tedious or complex tasks, if performed by humans, such as text extraction, automatic validation, document automation, or interoperability with external tools. Figure 5 suggests some of these tasks based on three key features: import from text or models, check the quality and publish or export. For instance, the import feature shall allow users to import an Excel file or an ad-hoc natural language text file containing the input data and produce the corresponding *SL files. The export feature shall transform an *SL file into other formats, namely Word, Excel, JSON, or Text. The check quality feature shall run syntactic and semantic-level rules to ensure that specifications are free of errors.

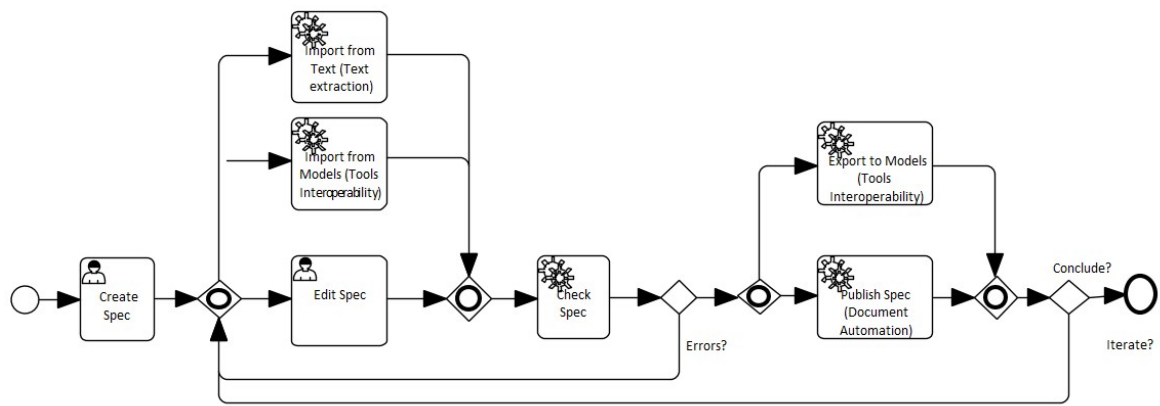

**Fig.5**. Simplified workflow of the ITLingo approach with automation tasks (BPMN notation).

These tools shall also include multiple transformations based on the *SL specifications. Figure 6 summarizes the key transformations that may be classified as T2M (Text-to-Model), M2M (Model-to-Model), and M2T (Model-to-Text) transformations.

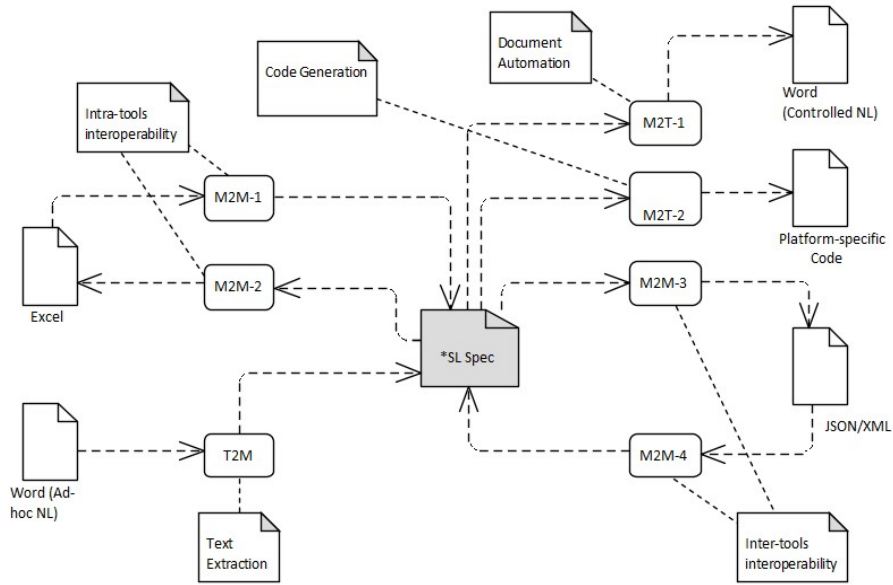

**Fig. 6.** Transformations that shall be supported by ITLingo tools (BPMN notation).

ITLingo-Studio deals with technical documents represented in multiple representations: ad-hoc and controlled natural language (NL) text, Excel, Word, and JSON or XML. We considered an Excel file a model since it is a tabular and highly structured representation. In contrast, we assumed that a Word file is like an NL text because it contains plain text with low-level formatting information. However, since it is not as structured as an Excel file, we considered it is a text and not a model (despite being internally organized in an archive of multiple XML files). Below, we describe each transformation type and then some implementation issues grouped by the technology used to support them. For instance, JSON and Text formats are generated using Xtend, while Word and Excel are generated using the Apache POI library.

**Text extraction (T2M Transformations).** Studio performs a T2M transformation during an ad-hoc NL text file import process. This transformation involves the execution of automatic text classification and extraction processes. The classification process identifies the sentences in the document provided and classifies them into a set of distinct categories as defined by the respective *SL language. The second process extracts relevant textual fragments from the original sentences into their equivalent representation in *SL. These processes have been implemented with artificial intelligence and natural language processing (NLP) frameworks using RapidMiner (an open-source platform for predictive analytics and data mining) or Stanford CoreNLP (a popular Java NLP framework). The implementation of automatic text extraction is a challenging task involving the integration and tuning of feature models, and it is still a work in progress.

**Interoperability with other tools (M2M Transformations).** M2M transformations are used during the import and export of a document in Studio. The import of an Excel file specifying a document (transformation M2M-1) generates its corresponding *SL file(s), depending on the file structure mode the user has selected. M2M-1 is implemented using the Apache POI library, simplifying the processing of Microsoft Office file formats. M2M-2 performs the reverse transformation (from *SL to Excel) and is implemented using the Apache POI library but uses an Excel template file. The remaining transformations export the *SL specification into JSON or XML formats (M2M-3) and vice-versa (M2M-4) to support interoperability with other software tools, like modeling tools or project management platforms.

**Document automation and code generation (M2T Transformations)**. The transformations from *SL into Word or similar formats (M2T-1) are based on document automation techniques with the Apache POI library and companion template files. We use Apache POI because it abstracts the complex XML structure that underlies Microsoft Office files. Additionally, we used template files to give a user more flexibility to customize the generated files' style and formatting. These templates have specific tags representing the dynamic or variability parts of the templates and identifying which properties should be placed there during the generation process. These tags are defined using the style (e.g., font type, size, colour) reflected in the generated file. Furthermore, some preliminary research has shown that is possible to implement transformations from ASL specifications into platform-specific code (M2T-2), for example, for low-code platforms (e.g., Quidgest Genio, Outsystems, Mendix) or full-stack platforms (e.g., Django, Node.js).

**Automatic validation**. Most of the work related to the quality of specifications traditionally depends on human-intensive tasks made by domain experts, which are time-consuming, error-prone, and have negative consequences on the success of the projects. We have explored automatic validation techniques to mitigate some of these limitations and increase the quality of such specifications concerning consistency, completeness, and unambiguousness. For instance, in [Sil2015, Car+2019] we discussed how to support such validations automatically: The consistency validation enforces that the information model underlying the *SL is well-formed or consistent under the *SL metamodel, which involves validations such as of attribute values, numeric sequences, or referential integrity. The completeness validation is based on the test's configuration resource that defines the level of completeness required for each *SL specification, for example, completeness at the model, viewpoint, or construct level. While inconsistencies and incompleteness can be automatically detected, ambiguities deal directly with the semantics of those elements, thus, harder to be detected by automatic processes. Still, some automatic validation can be applied, such as semantic analysis or terminology normalization.

### Case Studies, Projects, and Libraries

ITLingo languages and respective tools have been applied and validated in the scope of some academic case studies, as extensively discussed in the papers and theses published. However, some concrete projects were also used to validate these results. For example, in the scope of the RiverCure project (FCT, 2019-2022), we used RSL and ASL with the ITLingo-Studio to rigorously define the requirements of the RiverCurePortal, a complex web GIS platform for modeling, simulating and analyzing flood events. Furthermore, in the scope of the RiverCure project, we explored model-to-text transformations that generate code to Django, a high-level Python web framework. In the scope of the CIT-INOV project (2018-2021, ANI/FITEC), we researched the specification of reusable requirements in RSL for cross-cutting aspects such as cybersecurity and IT sustainability. We plan to continue exploring these aspects in our following projects like DikesFPro (2021-2023) and EV4EU (2022-2025).

In the scope of the ASI-CML-DMHU project (2019-2020, CM-Lisboa) we developed the EASL Excel template, which was successfully used by domain experts to rapidly elicit and represent the elements of the business and application layers of a large set of IT resources managed by the Municipal Department of Urban Hygiene of Lisbon City Council. The same approach with PSL and EASL Excel templates is being applied to projects like SIGA SNS (2021-2023, ACSS) and COFAP (2021-2022, INA).

We have also researched reusability features in these *SL languages, and thus, the definition of libraries of reusable specifications applied in multiple projects. For instance, for the RSL language, we defined libraries of requirement specifications for cross-cutting aspects such as usability [SS2020], privacy [FSG2020], and cybersecurity [GS2018].

# 3. Conclusion

Several surveys and studies have noticed that cost, and quality problems result from mistakes that occurred in the early phases of the projects, for instance: poor definition of the project vision and respective value for the organization; misalignment between IT and business resources; failures in project management practices; or poor and low-quality technical documentation. These facts have emphasized the need for improving socio-technical disciplines such as project management, enterprise architecture, requirements engineering, or system design. These studies also noted the need to reduce the efforts involved in traditional development processes, for example, by automating some human-intensive and error-prone tasks.

ITLingo is a research initiative that has proposed new languages, tools, and techniques to support users to improve such practices, mainly related to those disciplines. ITLingo users are IT engineers and managers in multiple roles like project managers, enterprise architects, business analysts, system architects, requirements engineers, or even system developers.

ITLingo results from our experience in previous research projects, projects developed jointly within the industry, and pedagogical experiences conducted in academic courses at IST, such as AMS (Systems Analysis and Modeling) or GPI (Information Systems Project Management).

ITLingo has gathered several students and researchers who have contributed with initial or genius solutions to some of the abovementioned problems and challenges. Figure 7 shows a glimpse of such contributions (concluded projects) around the languages and respective tools discussed above.

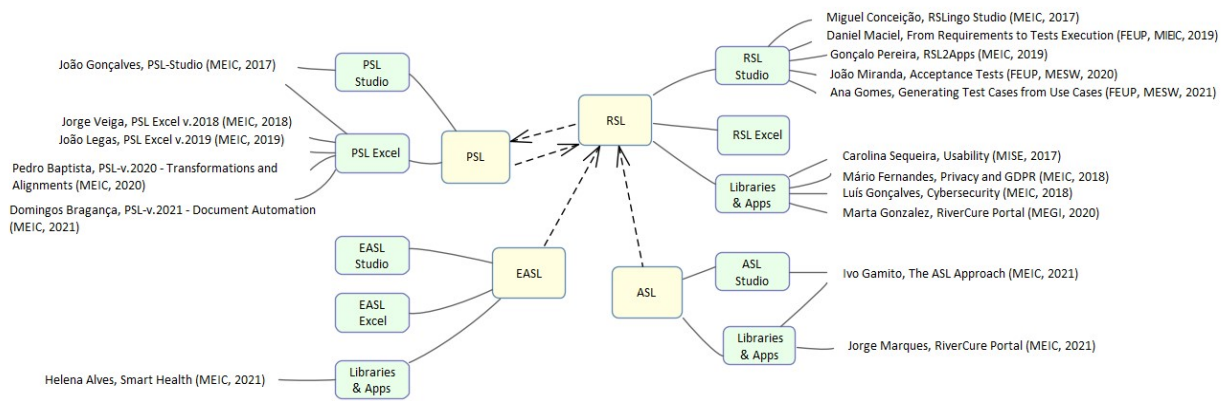

**Fig. 7**. Main contributions to the ITLingo initiative (Mindmap notation).

We plan to carry out activities to extend and foster the ITLingo initiative for future work. First, develop and deploy a multi-organization and multi-workspace platform (the ITLingo-Cloud), in which its users would easily set up their environment, easily manage workspaces and *SL-based documents, and get access to the discussed automation features, such as text extraction, advanced text validation, and model-based transformations. Second, continue producing and gathering reusable contents (specific to the concerned disciplines), both as reusable libraries of *SL specifications and *SL document templates. Third, extend the ASL language with new constructs that, for instance, would allow defining platform-independent specifications for robotic process automation, test automation, data science, or IoT-based applications. Fourth, research model-to-model and model-to-text transformations to explore the generation of software applications for low-code or no-code platforms. Fifth, progressively integrate some ITLingo results (e.g., Excel templates, ITLingo-Cloud platform, libraries of reusable specifications) into academic teaching processes.